# An mCherry biolaser based on microbubble cavity with ultra-low threshold

JIYANG MA[1,2,†], SHUOYING ZHAO[1,2,†], XUBIAO PENG[1,2], GAOSHANG LI[1,2], YUANJIN WANG[1,2], BO ZHANG[1,2,*] AND QING ZHAO[1,2,*]

[1]No. 5 Zhongguancun South Street, Center for Quantum Technology Research and Key Laboratory of Advanced Optoelectronic Quantum Architecture and Measurements (MOE), School of Physics, Beijing Institute of Technology, Beijing 100081, China
[2] No. 10 Xibeiwang East Street, Beijing Academy of Quantum Information Sciences, Beijing 100193, China
[†]These authors contributed equally to this work
* bozhang_quantum@bit.edu.cn, qzhaoyuping@bit.edu.cn

**Abstract:** Biolasers show considerable potential in the biomedical field. Fluorescent protein (FP) is a type of biomaterial with good luminescence efficiency that can be used as the luminescent gain medium in biolasers. Due to the higher cell/tissue permeability, lower cell phototoxicity, and relatively less background fluorescence than other fluorescent proteins, the red fluorescent protein is more suitable in biological applications. MCherry is the most extensively used high-quality red fluorescent protein because of its short maturation time and stable luminescence properties. In this study, using mCherry and microbubble cavity, we realize a highly stable mCherry fluorescent protein laser. The laser resonator achieves a quality factor of $10^8$, which is the highest Q factor among the currently available FP lasers. Moreover, this laser exhibits low threshold of 286 fJ, which can effectively protect the luminescent material from being damaged by pump light. Such a threshold is the lowest in the FP lasers as per our knowledge. The prepared laser shows excellent stability in a wide pH range with good photobleaching resistance and can be stored at 4◦C for nearly a month. Also, the laser can serve as a high-sensitivity molecular concentration detector with mCherry as biomarker, owing to its lasing threshold behavior.

## 1. Introduction

Bioluminescent materials [1-3] show high biocompatibility and biodegradability, including chlorophyll, riboflavin, and fluorescent protein (FP) etc. Lasers with biomaterials as a gain medium are referred to as biological lasers [4], which have recently attracted considerable attention among scholars. Unlike spontaneous emission, the laser generates stimulated radiation amplification of fluorescence through the interaction between the optical resonator and bioluminescent materials. Compared with biomaterial's fluorescence, biolasers offer high brightness, narrow linewidth, and high signal-to-noise ratio, which are more suitable for biosensing. Especially, the biological laser based on the whispering gallery mode microcavity with a high-quality factor and ultra-small mode volume facilitating strong interaction between the bioelement and optical mode can enhance the sensitivity [5,6]. This type of biological laser plays an essential role in cell tracking [7-9], biological detection [10-20], and biological imaging [21].

Fluorescent proteins (FPs) [22], first found in Aequorea victoria, is a nontoxic and biocompatible bioluminescent material. The absorption spectral of different mutants of FPs can cover the entire visible light band. Owing to these advantages, FPs have been instrumental in the biological field in the recent decades. Currently, because of the high quantum yield and popularity of green fluorescent protein (GFP), only green color FPs have been widely used to fabricate biological lasers [23-31]. However, due to the limited quality factor of the optical resonator, some FPs with low quantum yield in other wavebands, such as red fluorescent protein (RFP), have been scarcely reported in the production of biological lasers. In contrast with the green fluorescence, the red fluorescence has lower Rayleigh scattering, which is conducive to penetrating deeper tissues. In addition, the hemoglobin and water in biological tissues have low absorption coefficients in the band of red light. In summary, the red fluorescence shows higher cell tissue permeability, lower cell phototoxicity and relatively less background fluorescence, and hence can be used for deep tissue imaging in vivo. Therefore, FPs in the red wavelength regime are beneficial for imaging at different scales, from a single molecule to the entire organism. MCherry, mutated from the earliest RFP DsRed [32], is the most commonly used class of RFP with a short maturation time, good photobleaching resistance and excellent acid-base stability. It has been widely used in cell calibration and cell tracking [33-35].

Herein, we employ mCherry RFPs along with a high quality factor (Q factor) silica microbubble resonator [36] to fabricate an mCherry FP laser, as schematically shown in Figure 1(a). For the first time, pure mCherry solution is used as a luminescent gain medium to fabricate a FP laser whose lasing characteristics is intensively studied. Microbubble resonator is a whispering gallery type resonator with hollow structure, forming a microfluidic tunnel itself, and has been widely used in biosensing applications [37]. The laser cavity in this work achieves an ultrahigh Q factor of $10^8$, which is the highest Q factor among all the currently available FP lasers. Because of its high Q factor, the laser threshold is as low as 286 fJ, the lowest threshold in any type of FP lasers, as per our knowledge. At this threshold, the luminescent material can be effectively protected against damage from pump light. Due to the pre-eminent optical property of mCherry, the prepared laser shows good photobleaching resistance, great pH stability, and durability that can be preserved for nearly a month. Such high stability is of great importance in rigid environment biosensing, e.g., in solutions with varying pH and under long-time pumping excitation. The laser can serve as a high-sensitivity concentration detector with the red fluorescent protein as biomarker using its threshold behavior. Because the microcavity achieves a high Q factor, the minimum mCherry concentration to produce lasing reaches as low as 1.33 μM, which is the lowest concentration of the gain medium in the currently available FP lasers. The findings of this study will contribute to the understanding of luminescent materials for biological laser applications and guide trace biological detection, fluorescence resonance energy transfer (FRET), and other related fields.

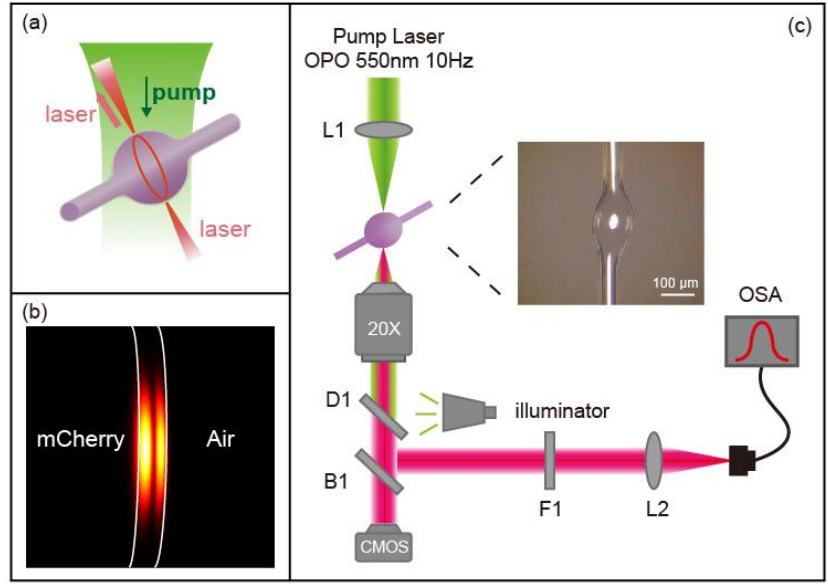

**Fig. 1.** Schematic of mCherry protein laser, optical mode simulation and experiment setup. (a) Schematic of mCherry laser; (b) Simulation of optical mode field with mCherry in the microbubble cavity; (c) Schematic of experimental setup: L1 and L2 are convex lenses, D1 is dichroic mirror, which can reflect green light and project red light, B1 is 50%/50% beam splitter, F1 is long-pass filter, which can passthrough red light and block green light, inset: microscopic image of the microbubble

## 2. Microbubble fabrication and experimental setup

The microbubble resonator is fabricated according to the following steps. We have used a silica capillary (Polymicro Technologies TSP100170) with an inner diameter, outer diameter, and sidewall thickness of 100 µm, 140 µm, and 20 µm, respectively. First, the capillary was immersed in a hot piranha solution (155◦C) to remove the thin polymer layer on its outer surface. Then, the capillary was immersed in 5% hydrofluoric acid to reduce its sidewall thickness from 20 µm to 10 µm. Then the capillary is drawn from both sides down to an outer diameter of 40 µm-50 µm under hydrogen flame heating. Next, the inside of the capillary was pressurized and blown into a microbubble cavity under carbon dioxide laser irradiation. Figure 1(b) shows the simulated optical mode distribution of the microbubble cavity filled with mCherry solution. It has a diameter and a wall thickness of 130 and 1.7 µm, respectively. Due to the ultrathin wall thickness of the resonator, the mode field shows a strong evanescent field inside the hollow area, which effectively interacts with mCherry to produce laser. The microscopic image of the microbubble is shown in the inset of Figure 1(c).

An experiment setup (Figure 1(c)) is used to excite mCherry in the microbubble cavity and collect the generated laser beam. The pulse duration of the pump light is 7 ns with a repetition rate and wavelength of 10 Hz and 550 nm, respectively. The pump light adjusts the energy and spot size through a polarizer and beam expander. Then, it passes through the convex lens with a focal length of 20 cm and finally irradiates the microbubble cavity with a spot size of 1-mm diameter. The laser is collected through a 20× objective lens with a numerical aperture (N.A.) of 0.4. A dichroic lens and long-pass filter are used to filter 550-nm pump light. Finally, an optical fiber is used to collect the red laser beam and transmit it to an optical spectrum analyzer (Princeton Instrument, HRS750, ProEM 512).

## 3. Calculation of the excitation and emission spectra of mCherry

According to fluorescence excitation mechanism of RFP [38], we calculate the absorption and fluorescence spectra of mCherry using quantum and molecular mechanics (QM/MM) method in the software Orca [39]. The detailed workflow is as follows. 1) The crystal structure of the protein mCherry was extracted from the protein database (PDBID: 2H5Q) [40]. 2) The protein was solvated in a cubic water box with an edge length of 15 Å, and the residue GLU215 was protonated. In the QM/MM simulation, the chromophore (the ball-and-stick parts, Figure 2 (a)) was selected as the QM region, and others were selected as the MM regions. The hydrogen atoms in the chromophore were added using the software GaussView, and hydroxyl ion (OH-) and hydrogen ion (H+) were added at the QM/MM interface of the chromophore to complement the effect of covalent bond breaking. The algorithm L-BFGS is used to pre-optimize the entire system under the molecular force field of CHARMM 27. 3) After the above preparation, we further optimized the structures of the ground and excited states, where the optimization level is WB97X-D3/cc-PVDZ [41,42]. In the calculation process, the chromophore is set as the center, and the atoms within 12 Å are the active region. 4) After structural optimization, we use the double hybrid function DSD-PBEP86 [43], with higher precision, to calculate the single point energy for excitation. In the calculation, the D3 correction and higher-level basis aug-cc-pVTZ were used [42].

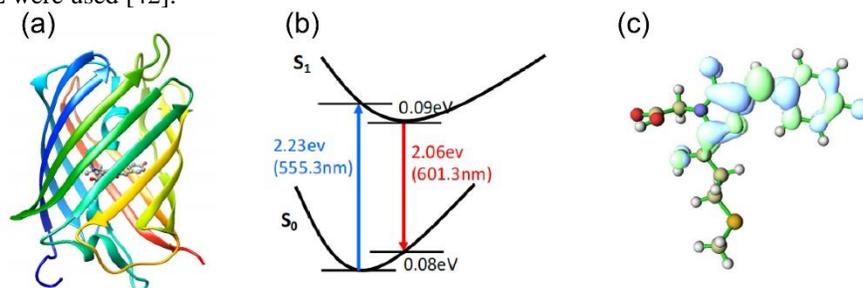

**Fig. 2.** Three-dimensional structure, energy level, and electron transfer during the fluorescence of mCherry. (a) Three-dimensional structure diagram of mCherry. (b) Energy levels for vertical excitation and fluorescence emission calculated using the QM/MM method. (c) The distribution of electron–hole at the beginning of the fluorescence process. The green and blue regions represent electron and hole densities, respectively.

Figure. 2 (b) shows the obtained energy levels for vertical excitation. The absorption and fluorescence emission peaks of mCherry are at 555.3 and 601 nm, respectively, which are consistent with the experimental results reported in a previous investigation [44] but slightly blue-shifted. According to the calculation results, we use the 550-nm laser as the pump light to generate the red laser. Finally, we calculate the electron–hole distribution along the chromophore in mCherry at the beginning of fluorescence emission using the wave function analysis software Multiwfn [45,46] (Figure. 2 (c)). During the transition from the excited to ground state, the electron moves from right to left, i.e., moving away from the aromatic ring of the chromophore.

## 4. Results and Discussion

In this section, we characterize the fabricated mCherry FP laser from the pumping threshold, luminescence stability, pH stability and photobleaching resistance, respectively. In the end, we discuss the application of mCherry FP laser as a protein concentration sensing meter.

### 4.1 Device characterization–Spectrum and threshold measurement

We prepare a microbubble cavity with a diameter and wall thickness of 130 and 1.7 μm, respectively (shown in the inset of Figure 1(c)). Subsequently, the mCherry solution with a concentration of 39.7 µM is injected into the microbubble cavity. The evanescent mode field inside the microbubble can interact with the mCherry solution and produce a laser. Figure 3(a) shows the laser emission spectra of mCherry protein with a concentration of 39.7 µM excited under different pump energies, with the inset in the upper left corner demonstrating the microbubble image excited by the pump light using a CMOS camera. This confirms the simulation results that the optical mode distributes on the side wall of the microcavity, where the gain medium interacts with the microcavity and produces a laser. The lasing spectra are collected with a grating of 1200 g/mm in the optical spectrum analyzer. Unlike the broadband fluorescence emission, the laser shows distinct sharp peaks in the lasing spectra. The maximum lasing emission is at the wavelength of 636.3 nm which shows a redshift compared with the calculated maximum fluorescence emission peak of mCherry (601 nm). This is because fluorescence is a spontaneous emission process, whereas laser belongs to the stimulated emission. According to a previous study [47], to obtain the stimulated emission spectrum, one needs to multiply the spontaneous emission spectrum with a factor of $\lambda^4$. The inset in the lower right corner is the spectrum collected using a high-resolution grating of 2400 g/mm. The obtained lasing linewidth is 0.023 nm, which is limited by the resolution of the spectrum analyzer. The actual linewidth should be less than $6.4\times10^{-6}$ nm, which depends on the Q factor of the cavity, as shown in the inset of Figure 3(b). The linewidth of the optical mode is 3.5 MHz at 780 nm wavelength, corresponding to the Q factor of $10^8$. Due to the high Q factor, the lasing threshold (Figure 3(b)) reaches as low as 1.15 µJ/mm$^2$. The simulation result (Figure 1(b)) shows that the mode volume of the evanescent field interacting with mCherry solution is only 22 µm$^3$, absorbing merely 286 fJ at laser threshold. This is the lowest threshold among all currently available FP lasers. A low threshold is critical in the application of biolasers, which can ensure that the bioluminescent materials are resistant to damage caused by pump light and avoid the introduction of other effects in the biological detection processes, such as the thermal effect and nonlinear effect of microcavity's material.

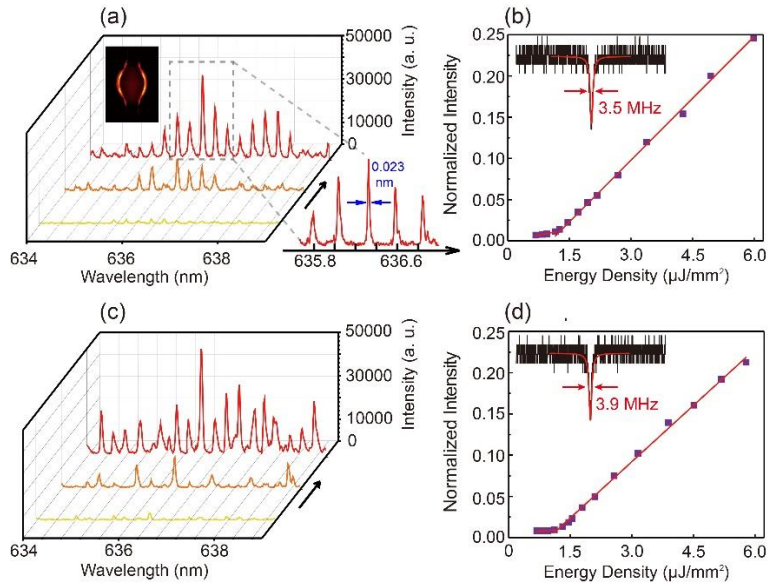

**Fig. 3.** Lasing spectra and threshold measurement. (a) Spectra of the mCherry laser under different pump energies. The arrow indicates the pump energy increment, inset: upper left corner: microbubble image excited by the pump light, lower right corner: spectrum collected using a high-resolution grating of 2400 g/mm. (b) Threshold measurement of the mCherry laser, inset: Lorentz transmission spectrum of the microbubble cavity at 780 nm, corresponding to a Q factor of $10^8$. (c) Spectra of the mCherry laser under different pump energies after nearly a month. The arrow indicates the pump energy increment. (d) Threshold measurement of the mCherry laser nearly a month later, inset: Lorentz transmission spectrum of the microbubble cavity at 780 nm, corresponding to a Q factor of $9.8 \times 10^7$.

*4.2 Laser Stability verification*

### 4.2.1 Luminescence stability verification

The laser exhibits good luminescence stability. We seal both ends of the microbubble cavity with ultraviolet glue to ensure that the mCherry is not polluted or evaporated and store it in the sample box to prevent microbubble from contamination. When not used, the sample is stored in a refrigerator at 4◦C. After nearly a month, the luminescence properties of the mCherry protein laser changes slightly while still maintaining the low threshold characteristics. Figure 3 (c) and (d) shows laser emission spectra as well as the threshold measurement and the Lorentz transmission line of the optical mode after nearly a month, respectively. Compared with the data shown in Figure 3 (a) and (b), no obvious change is observed in the Q factor and spectra of the microcavity laser. The laser threshold remains as low as 1.16 µJ/mm$^2$. In fact, the preservation time can last even longer, which is not limited to one month's test. The long-term reusability of our mCherry laser is of great significance in practical applications.

### 4.2.2 Stability on pH value

Next, we test the pH stability of the mCherry laser. We change the pH value of the mCherry FP solution from 3 to 12 at a consistent concentration. The experimental results show that the mCherry FP laser has high pH stability. Figure 4 (a) shows the lasing spectra of different pH values under a pump excitation of 29.7 µJ/mm$^2$. The lasing behavior shows good stability with the variation of pH value. Figure 4(b) shows the laser threshold at different pH values. The threshold varies slightly in the pH range between 3 and 12, presenting good consistency. These results indicate that mCherry protein laser bears high stability in a wide pH range, offering considerable potential in the biological detection of aqueous media with different pH levels.

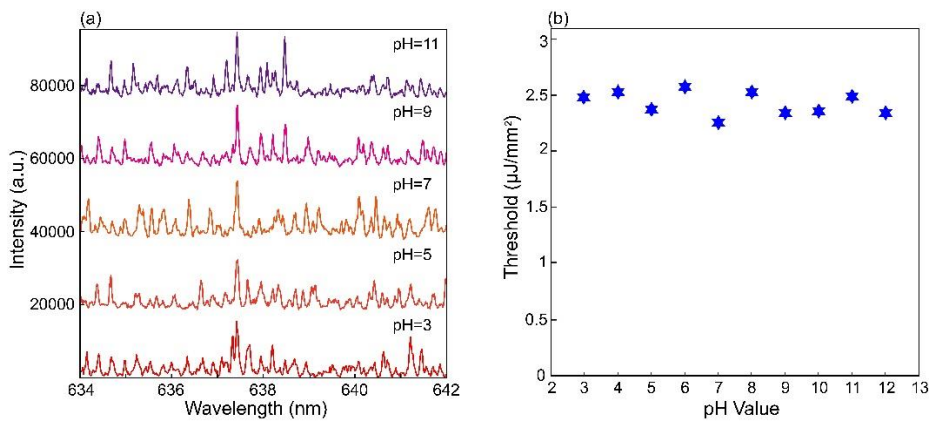

**Fig. 4.** Lasing spectra and threshold under different pH values. (a) Lasing spectra with different pH values; (b) Threshold values with different pH values

### 4.2.3 Stability on photobleaching resistance

The laser also shows good photobleaching resistance. An mCherry laser is continuously excited using the pump light at 5.18 µJ/mm$^2$ with 10 Hz repetition rate. The laser spectra after 30 mins excitations are presented in Figure 5(a). Figure. 5(b) shows the evolution of the normalized intensity of the peak laser line of mCherry at different times. The laser intensity decreases to 1/e of the initial value in the first 15 mins and retains 20 % of the intensity in 30$^{th}$ minute. By comparison, Figure 5(c) shows the laser generated at the same concentration of the GFP. However, the photobleaching effect of the GFP laser is severe whose laser intensity decreased by >90% within 2 mins of excitation. Figure. 5(d) shows the evolution of the normalized intensity of the peak laser line of GFP. Evidently, mCherry protein laser has strong optical stability after long-time irradiation of pump light. The good photobleaching resistance makes mCherry laser have great practical application prospects over the current GFP lasers.

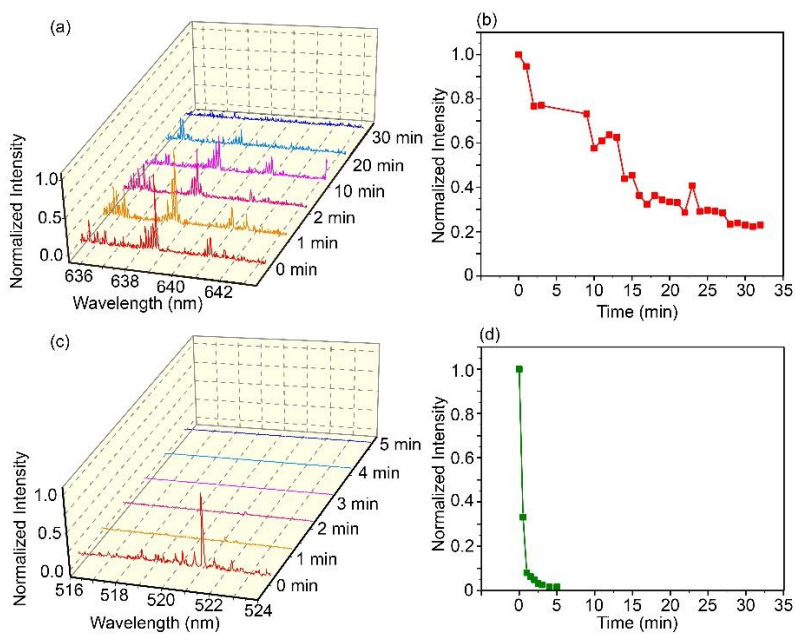

**Fig. 5.** Comparison of the photobleaching resistance between mCherry and GFP laser. (a) Normalized spectra of mCherry laser within 30 minutes under continuous excitation of pulsed pump light; (b) Normalized intensity of the mCherry peak laser in different times; (c) Normalized spectra of GFP laser within 5 minutes under the continuous excitation of pulsed pump light; (d) Normalized intensity of the GFP peak laser in different times

### 4.3 The application as a protein concentration sensing meter

Finally, we realize high-sensitivity mCherry protein concentration detection based on the laser threshold characteristics. Figure 6(a) shows the laser thresholds in different concentration mCherry solutions. The threshold value decreases as the mCherry concentration increases from 1.33 to 39.70 µM. The detection sensitivity is rather high at a low concentration regime. For example, the threshold values are 34.11, 19.7, and 5.96 µJ/mm$^2$ at concentrations of 1.33, 1.6, and 2 µM, respectively, corresponding to a sensitivity of 42.01 µJ/mm$^2$/µM. Figure 6(b) shows four typical threshold measurements at low mCherry concentrations of 1.6 and 2 µM, medium concentration of 4 µM, and high concentration of 10 µM. Furthermore, the concentration can be detected by testing the lasing intensity at a fixed pumping energy. The lasing spectra of four mCherry concentrations under the same pump energy of 46.71 µJ/mm$^2$ are presented in Figure 6(c). Consequently, the laser intensity decreases with the decrease in mCherry concentrations. In summary, the mCherry concentration can be detected by testing the laser threshold or intensity, and the sensitivity can reach up to 42.01 µJ/mm$^2$/µM at a low concentration regime. This preliminarily confirms that our laser can be utilized as a molecular concentration detector by coupling with mCherry protein molecules as biomarker.

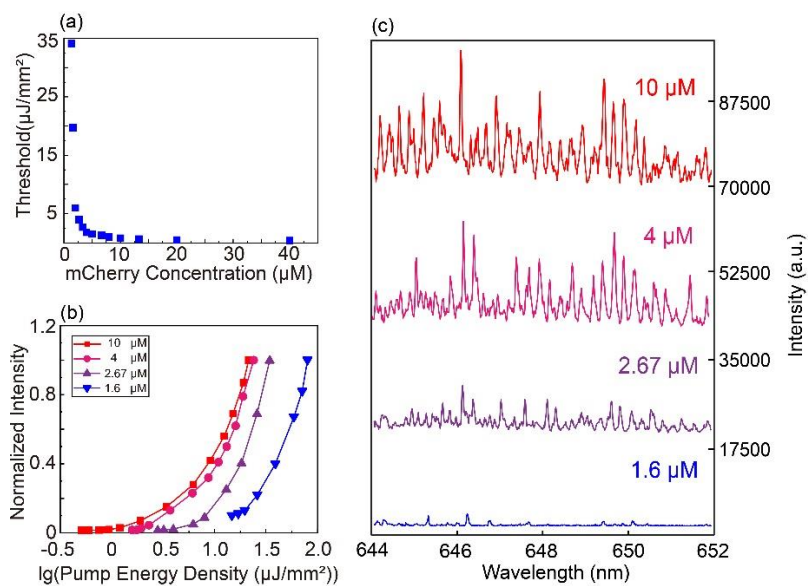

**Fig. 6.** Mcherry concentration detection. (a) Laser threshold variation with mCherry concentration. (b) Laser threshold measurement with the mCherry concentration of 1.6, 2.67, 4, and 10 µM. (c) Laser spectra with different mCherry concentrations under the same pump energy density of 46.71 µJ/mm$^2$

## 5. Conclusions

In summary, an mCherry FP laser is fabricated using mCherry and silica microbubble resonator. The Q factor of the prepared laser cavity reaches $10^8$, which is the highest among the currently available FP lasers; the corresponding laser threshold is as low as 286 fJ, the lowest in any type of FP lasers. Owing to the excellent characteristics of mCherry, the proposed laser exhibited good stability. It can be stored at 4◦C for nearly a month, and has excellent pH stability ranging 3–12. Furthermore, due to the strong photobleaching resistance of mCherry, the laser can function for a long time under the action of pump light. Red fluorescence could be easily observed in human tissues compared with green fluorescence because of the low absorption rate of red light in human tissues. Therefore, the laser can be used for the biological perception and observation of relevant life activities in deep tissue. Based on its low threshold characteristics, The laser can serve as a high-sensitivity molecular concentration detector with the red fluorescent protein as biomarker whose detection limit can reach as low as 1.33 µM and the sensitivity of 42.01 µJ/mm$^2$/µM at low concentration regime. In future, we plan to use the laser for the detection of trace biomolecules and study the FRET: By connecting the GFP and mCherry to both termini of the protein to be detected and interacting with the microbubble resonator, the FRET laser can be generated under the excitation of pump light. The conformational change of the target protein leads to the change in the distance between GFP and mCherry, thus affecting the efficiency of FRET. Therefore, the conformational change of the protein can be studied by monitoring the laser intensity and threshold, so as to realize the study on spatiotemporal regulation and function of proteins [48].

**Acknowledgement** The research was supported by China National Postdoctral Program for Innovative Talents (BX20200057).

**Disclosures.** The authors declare no conflicts of interest.